# Chirp Localization via Fine-Tuned Transformer Model:
# A Proof-of-Concept Study

Nooshin Bahador[1], Milad Lankarany[1,2]

1: Krembil Research Institute, University Health Network, Toronto, Canada.
2: Institute of Biomedical Engineering and Department of Physiology, University of Toronto, Toronto, Canada.

- [Access to Trained Vision Transformer on Synthetic Spectrograms](#)
- [Download Link for 100,000 Synthetic Spectrogram Dataset for Chirp Localization (Images & Corresponding Labels)](#)
- [Repository with PyTorch Implementation for Fine-Tuning the Vision Transformer](#)
- [Python package for generating synthetic chirp spectrogram (Images & Corresponding labels)](#)

**Abstract**

Spectrograms are pivotal in time-frequency signal analysis, widely used in audio processing and computational neuroscience. Chirp-like patterns in electroencephalogram (EEG) spectrograms—marked by linear/exponential frequency sweep—are key biomarkers for seizure dynamics, but automated tools for their detection, localization, and feature extraction are lacking. This study bridges this gap by fine-tuning a Vision Transformer (ViT) model on synthetic spectrograms, augmented with Low-Rank Adaptation (LoRA) to boost adaptability. We generated 100,000 synthetic spectrograms with chirp parameters, creating the first large-scale benchmark for chirp localization. These spectrograms mimic neural chirps using linear/exponential frequency sweep, Gaussian noise, and smoothing. A ViT model, adapted for regression, predicted chirp parameters. LoRA fine-tuned the attention layers, enabling efficient updates to the pre-trained backbone. Training used MSE loss and the AdamW optimizer, with a learning rate scheduler and early stopping to curb overfitting. Only three features were targeted: Chirp Start Time (Onset Time), Chirp Start Frequency (Onset Frequency), and Chirp End Frequency (Offset Frequency). Performance was evaluated via Pearson correlation between predicted and actual labels. Results showed strong alignment: 0.9841 correlation for chirp start time, with stable inference times (~137–140s) and minimal bias in error distributions. This approach offers a tool for chirp analysis in EEG time-frequency representation, filling a critical methodological void.

## 1. Introduction

Spectrogram analysis has become a cornerstone in various fields, from computational neuroscience to audio signal processing and other biomedical applications. Spectrograms, which represent signals in the time-frequency domain, are widely used due to their ability to capture intricate spectral details and high-level semantic information. This has led to their extensive adoption in tasks such as automatic speech recognition (ASR), pathological speech detection, and audio synthesis. For instance, numerous automatic pathological speech detection methods rely on spectrogram input representations, combining self-supervised learning embeddings with Short-Time Fourier Transform (STFT) spectrogram features to refine speech signals (Kaloga et al., 2024; Hung et al., 2022; Zheng et al., 2024). Similarly, multi-band spectrogram-based approaches have been employed to isolate individual sound sources from complex audio mixtures (Wang et al., 2023).

The transformation of signals into spectrograms using methods such as STFT, Constant-Q Transform (CQT), and Wavelet Transform (WT), combined with filters like Mel, Gammatone, and linear filters, has enabled the development of robust deep learning models for various applications (Pham et al., 2024; Yao et al., 2024). Generative models, such as those using diffusion processes and autoencoders, have also leveraged Mel-spectrogram representations for tasks like audio generation and speaker conversion (Zhao et al., 2025; Hou et al., 2024; Götz et al., 2024). Furthermore, spectrograms have been integral to text-to-speech systems, where outputs are transformed into spectrograms within the spectral domain (Lou et al., 2024; Zhang et al., 2024).

Despite the advancements in audio signal processing, the application of spectrogram-based analysis using deep models in other domains, such as EEG data analysis, remains limited. EEG datasets, in contrast to audio data, are typically smaller and exhibit significant variability in characteristics, posing challenges for model training and generalization (Jiang et al., 2024). However, recent efforts, such as the development of Large Brain Models, aim to address these challenges by learning generic representations from EEG data (Jiang et al., 2024). This highlights the need for further research to bridge the gap between spectrogram-based approaches in audio and neural domains.

In seizure monitoring, spectrograms are very important because they display how the frequency of brain activity changes over time. These changes help identify patterns linked to seizures. One of these patterns is called a "chirp". A "chirp-like" pattern in EEG/LFP spectrograms (recordings of brain activity) shows a steady increase or decrease in frequency within a narrow band, often appearing as linear or curved lines. This pattern can occur during the start, middle, or end of a seizure and is linked to the epileptogenic zone (the brain area where seizures begin). Chirps are important because they act as biomarkers, helping researchers understand seizure mechanisms, track their progression, and evaluate the effects of treatments. By studying these patterns, scientists can improve seizure detection, monitor treatment effectiveness, and advance epilepsy research (Bahador et al., 2024).

Currently, there is no trained model capable of automatically identifying chirp patterns in time-frequency representations, accurately localizing them, and extracting their features. This paper aims to address this gap by developing a robust and automated approach for chirp detection, localization, and feature extraction, thereby advancing the chirp-based seizure monitoring systems.

## 2. Method

The study employed an automated pipeline for data generation, model training, and evaluation, leveraging a Vision Transformer (ViT) architecture adapted for regression tasks. The data generation process involves creating spectrograms from linear and exponential chirp signals, adding Gaussian noise, and applying smoothing filters. The generated spectrograms are preprocessed and split into training and testing sets, with labels normalized for model training. The ViT model processes input images by dividing them into patches, encoding them through a transformer encoder, and predicting regression outputs via a regression head. The model is optimized using the AdamW optimizer with a learning rate scheduler and evaluated using metrics such as Pearson correlation and inference speed. Detailed steps, mathematical formulations, and architectural specifics, including the application of LoRA (Low-Rank Adaptation) for fine-tuning, are provided in Appendix A for reproducibility and further reference.

The flowchart (Figure 1) outlines the steps for generating synthetic spectrograms: (1) Initialize a zero spectrogram matrix; (2) Generate chirp signals, either linear or exponential; (3) Map chirp to spectrogram using time bins, frequency bins, and Gaussian weights; (4) Add noise and smooth the spectrogram; (5) Save the spectrogram as an image and store chirp parameters in a CSV file. The details of each step are explained in the following subsections.

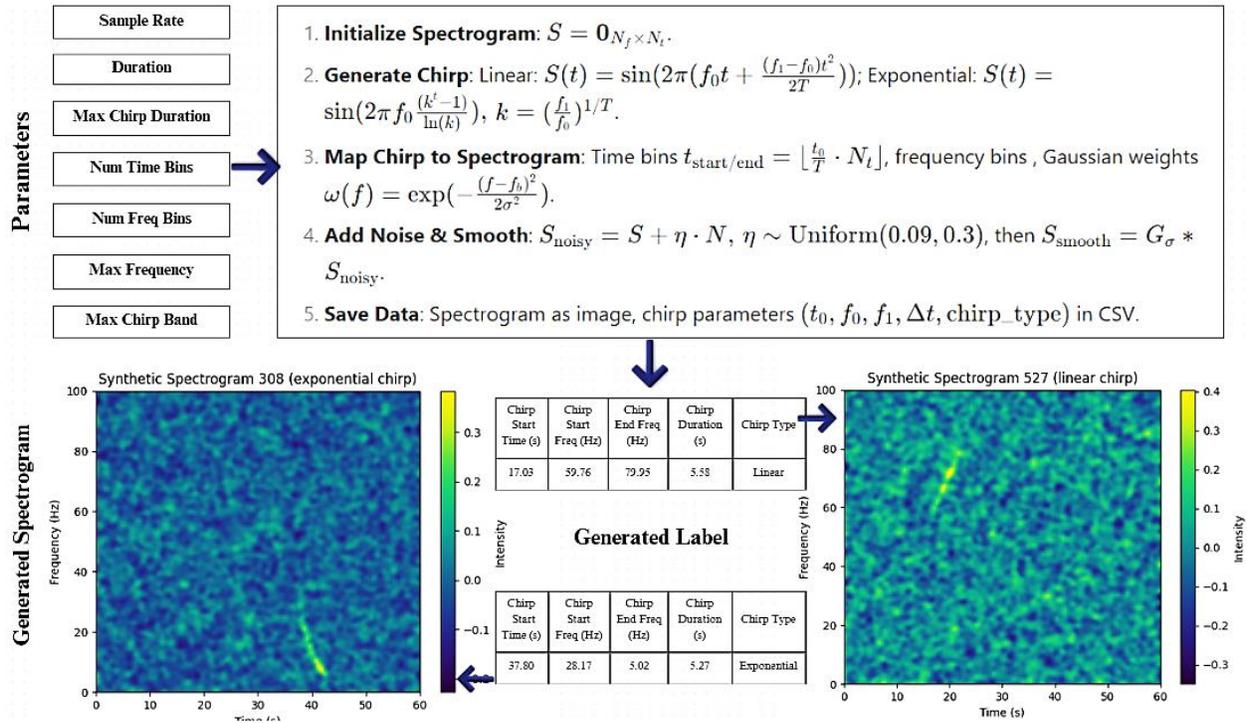

**Figure 1:** Flowchart illustrating the process of generating synthetic spectrograms with embedded chirp signals and their corresponding labels, including initialization, chirp generation, mapping to spectrogram, noise addition, smoothing, and data saving.

The pipeline (Figure 2) begins with data preparation, including label normalization, dataset splitting, and image preprocessing, followed by a regression task where images are processed through patch embedding, a transformer encoder, and a regression head to predict outputs. The model is optimized using MSE loss and AdamW, leveraging a Vision Transformer (ViT) backbone for robust feature extraction and regression. The details of each step are provided in the following subsections.

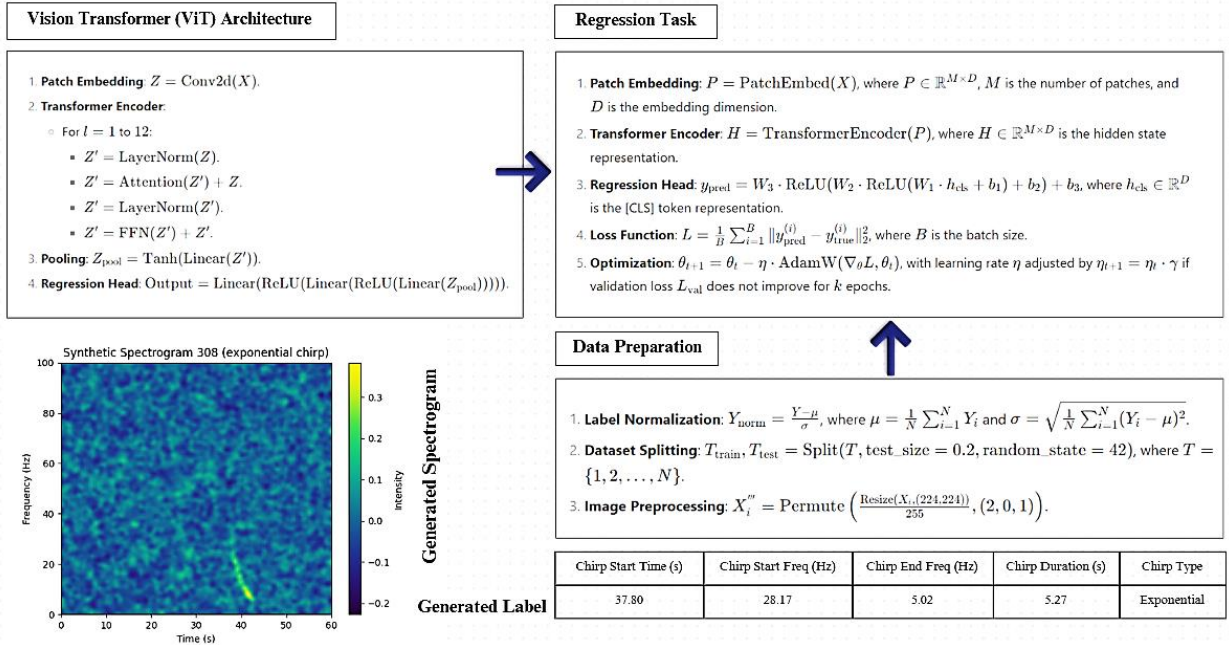

**Figure 2:** Pipeline for Regression Using Vision Transformer (ViT): From Data Preparation to Model Optimization

## 3. Results

The consistent decrease in training loss (Figure 3) suggests that the model is effectively minimizing its error on the training dataset. The steep decline in the initial epochs indicates rapid learning, while the gradual reduction in later epochs suggests convergence toward an optimal solution.

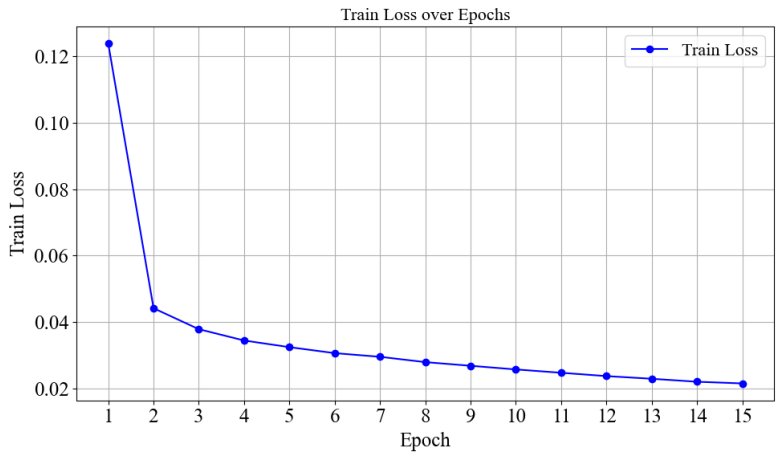

**Figure 3: Train Loss over Epochs:** The figure shows the training loss over 15 epochs. The loss decreases steadily from 0.1237 at epoch 1 to 0.0216 at epoch 15, indicating that the model is learning and improving its performance on the training data.

The test loss (Figure 4) steadily decreases until epoch 11. After epoch 11, it starts to slightly increase, which triggers early stopping.

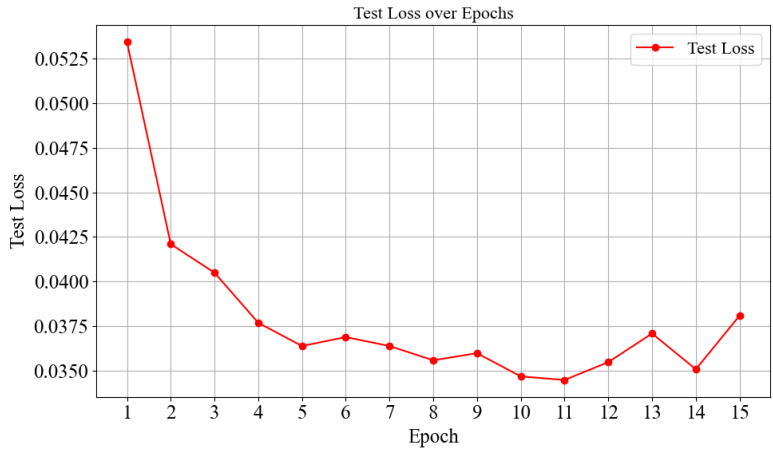

**Figure 4: Test Loss over Epochs:** The figure illustrates the test loss over 15 epochs. The test loss fluctuates slightly but remains relatively stable, ranging between 0.0347 and 0.0381

The increasing trend in correlation values (Figure 5) indicates that the model is improving its ability to capture the relationships between the predicted and actual values for the metrics. The high correlation values (close to 1) suggest strong agreement between the model's predictions and the ground truth. The slight variations in later epochs may reflect minor overfitting or noise in the data.

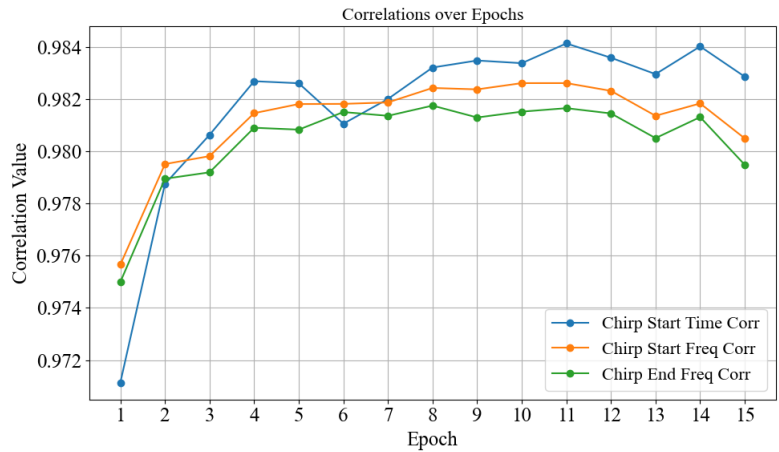

**Figure 5: Correlations over Epochs:** The figure displays the correlation values for three metrics—Chirp Start Time, Chirp Start Frequency, and Chirp End Frequency—over 15 epochs. All three correlations show a gradual increase, with Chirp Start Time Correlation reaching the highest value (0.9841) at epoch 11.

The stability in inference time (Figure 6) suggests that the computational cost of making predictions does not change significantly as the model trains.

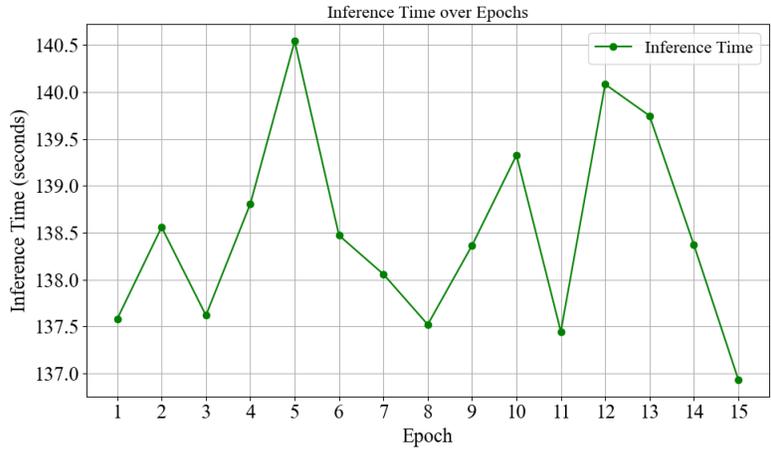

**Figure 6: Inference Time over Epochs:** The figure presents the inference time over 15 epochs. The inference time remains relatively stable, ranging between 136.9369 and 140.5395 seconds, with no significant trend.

Figure 7 illustrates the error distribution for all data, including both training and test sets, across three variables: Chirp Start Time, Chirp Start Frequency, and Chirp End Frequency. The majority of prediction errors are small, approximately $10^0$. The error distributions are fairly symmetric and show no significant skew, indicating that the model's predictions do not exhibit systematic bias. Additionally, the absence of long tails in these distributions suggests that the model rarely produces extreme outliers or large errors.

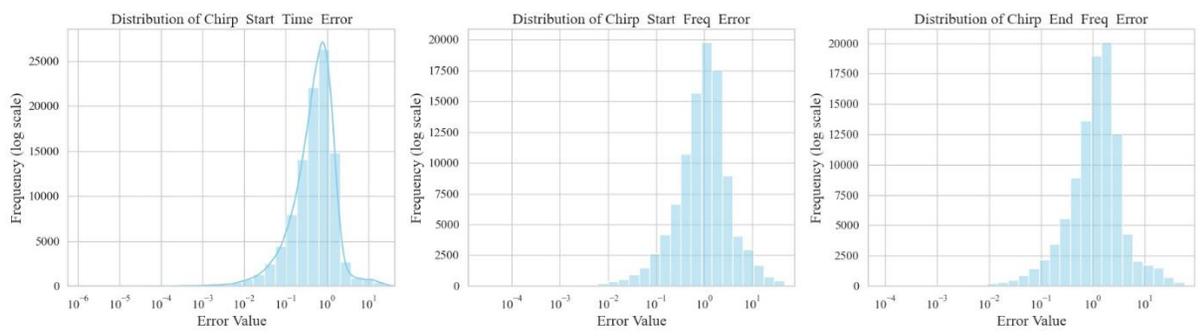

**Figure 7: Distribution of Prediction Errors for Training and Test Data:** The figure shows the histograms of prediction errors for three variables—Chirp Start Time, Chirp Start Frequency, and Chirp End Frequency—on both training and test datasets. The errors are plotted on a logarithmic scale to better visualize the distribution of smaller error values. Each subplot represents the error distribution for one variable, with the x-axis showing the error values and the y-axis representing the frequency of occurrence on a log scale.

As seen in the terminal output (Figure 8), the predicted chirp pattern for 5 samples shows a close approximation to the real label, with minor errors.

```
Predictions and Real Labels:
Sample 1:
  Predicted: A chirp pattern was observed starting at time 44.91 with a start frequency of 26.78 Hz and an end frequency of 23.65 Hz.
  Real: A chirp pattern was observed starting at time 43.33 with a start frequency of 24.33 Hz and an end frequency of 19.14 Hz.
---
Sample 2:
  Predicted: A chirp pattern was observed starting at time 12.42 with a start frequency of 62.61 Hz and an end frequency of 38.40 Hz.
  Real: A chirp pattern was observed starting at time 13.08 with a start frequency of 63.83 Hz and an end frequency of 39.03 Hz.
---
Sample 3:
  Predicted: A chirp pattern was observed starting at time 11.92 with a start frequency of 53.92 Hz and an end frequency of 32.47 Hz.
  Real: A chirp pattern was observed starting at time 13.58 with a start frequency of 54.79 Hz and an end frequency of 29.66 Hz.
---
Sample 4:
  Predicted: A chirp pattern was observed starting at time 6.72 with a start frequency of 94.16 Hz and an end frequency of 98.25 Hz.
  Real: A chirp pattern was observed starting at time 7.35 with a start frequency of 94.70 Hz and an end frequency of 97.22 Hz.
---
Sample 5:
  Predicted: A chirp pattern was observed starting at time 36.20 with a start frequency of 97.70 Hz and an end frequency of 99.22 Hz.
  Real: A chirp pattern was observed starting at time 38.27 with a start frequency of 99.59 Hz and an end frequency of 100.00 Hz.
---
```

Figure 8: Model Output and Ground Truth Comparison

## 4. Discussion

### 4.1. Advancements in spectrogram-based methods

Spectrogram-based approaches have revolutionized signal processing, particularly in audio and biomedical domains, by enabling the extraction of intricate spectral and temporal features. In audio signal processing, the integration of self-supervised learning embeddings with Short-Time Fourier Transform (STFT) spectrograms has enhanced tasks such as pathological speech detection (Zheng et al., 2024). Multi-band spectrogram techniques have further improved the isolation of individual sound sources in complex mixtures (Wang et al., 2023), while generative models like diffusion processes and autoencoders have demonstrated exceptional capabilities in audio generation and speaker conversion using Mel-spectrograms (Zhao et al., 2025; Hou et al., 2024; Götz et al., 2024). The integration of spectrograms into text-to-speech systems has further expanded their utility, enabling the transformation of outputs into spectrograms within the spectral domain (Lou et al., 2024; Zhang et al., 2024). Moreover, the use of spectrograms in multimodal decoders, where vision, text, and speech are embedded into a shared subspace of a transformer model, has facilitated the development of generative models for discretized mel-spectrograms (Gupta et al., 2024). These advancements highlight the versatility of spectrograms in capturing both high-level semantic information and fine-grained spectral details. However, challenges persist in spectrogram analysis, particularly in maintaining semantic coherence and preserving detailed visual information during generative reconstruction tasks (R et al., 2025). Additionally, the domain gap between synthetic and real spectrograms limits the generalizability of models trained on synthetic data (Liang et al., 2025). Bridging this gap remains a critical area for improvement.

While spectrogram techniques have been extensively applied to audio signals, their use in EEG data analysis is still nascent. EEG datasets are typically smaller and exhibit significant variability, posing challenges for robust model development (Jiang et al., 2024). Recent efforts, such as the development of Large Brain Models, aim to address these limitations by learning generic representations from EEG data, offering a promising direction for future research (Jiang et al., 2024).

Fine-tuning and representation learning have played pivotal roles in advancing spectrogram-based techniques. For instance, deep learned representations extracted from spectrograms using CNNs and ResNet architectures have been fine-tuned for specific tasks (Mallol-Ragolta et al., 2021). Joint-Embedding Predictive Architectures (JEPA) have further enhanced representation learning by predicting target representations from context-based spectrogram components (Riou et al., 2024). Conditional learning and impact-aware training objectives have also been employed to generate new data samples and identify root causes effectively (Kataria et al., 2022; Ouyang et al., 2025).

In-context learning (ICL) and uncertainty quantification have further expanded the applicability of spectrogram-based models. ICL, combined with retrieval-augmentation techniques, has improved segmentation tasks by leveraging ground truth annotations (Gu et al., 2025; Cosarinsky et al., 2025). Gaussian Processes (GPs) have been utilized to quantify uncertainty in regression and classification tasks, enhancing model reliability (Huang et al., 2025).

Visual interpretability and vector quantization have also contributed to the field. Techniques like Grad-CAM and Eigen-CAM have enabled cross-modal interpretability by highlighting disease-affected regions in spectrograms (Jahin et al., 2025). Visual vector quantization has improved cross-modal understanding and generation quality by leveraging discrete representations (Zhang et al., 2025).

Innovative architectures, such as Attentive Time-Frequency Neural Networks (ATFNN) and Spatial Audio Spectrogram Transformers, have demonstrated strong performance in tasks like speech emotion recognition and sound event detection (Lu et al., 2022). The Spatial Audio Spectrogram Transformer demonstrated strong performance in sound event detection, spatial localization, and distance estimation (Zheng et al., 2024).

Pre-trained text-to-spectrogram diffusion models and vision models have further expanded the utility of spectrograms by operating within shared latent spaces (Chen et al., 2024; Zhan et al., 2025). Transfer learning from computer vision models, such as ResNet and EfficientNet, has significantly improved audio detection tasks using spectrogram features (Pham et al., 2024). Large foundational vision models were also applied to spectrograms for audio processing, making predictions within a latent space (Fei et al., 2023). Vision Language Models (VLMs) were used as classifiers for audio spectrograms in a few-shot learning setup. These models were able to recognize the content in audio recordings by being provided with corresponding spectrogram images. In this few-shot approach, VLMs were given example spectrograms for each class and prompted to classify a new spectrogram image based on those examples (Dixit et al., 2024).

Spectrograms served as inputs for large language models as well, contributing to the improvement of Spoken Question Answering and Speech Continuation tasks (Nachmani et al., 2023). Vector Quantized Generative Adversarial Network is a generative model that uses vector quantization (VQ) to map high-dimensional data into a discrete latent space. By "compact sampling space," the model is designed to represent data in a lower-dimensional latent space. The model incorporates a new type of loss function that is based on spectrograms. Perceptual loss is computed in the spectrogram domain. This loss ensures that the generated samples are not only mathematically close to the target but also perceptually realistic for frequency-sensitive data (Iashin et al., 2021).

### 4.2. Challenges and Future Directions

Models were fine-tuned for various downstream tasks. However, in contrast to text data, EEG datasets were typically smaller in volume and exhibited significant variability in format. The development of Large Brain Model for learning generic representations from EEG data was explored to address these challenges (Jiang et al., 2024). Key challenges in generative reconstruction involved the need to maintain semantic coherence, manage complexity, and preserve detailed visual information (R et al., 2025). There might be a domain gap between synthetic spectrograms and real spectrograms (Liang et al., 2025).

Future research should focus on addressing the challenges in spectrogram analysis, particularly in maintaining semantic coherence and bridging the gap between synthetic and real spectrograms. Additionally, further exploration of spectrogram-based approaches in EEG data is needed to unlock their potential in biomedical applications. The development of large-scale models for EEG data, such as the Large Brain Model, represents a significant step forward in this direction (Jiang et al., 2024). By leveraging the advancements in audio signal processing and extending them to EEG data, researchers can unlock new possibilities for spectrogram analysis in diverse domains.

## 5. Acknowledgments

The author thanks the Canadian Neuroanalytics Scholars (CNS) Program for supporting this research.

**References**

Bahador, N., Skinner, F., Zhang, L., & Lankarany, M. (2024). Ictal-related chirp as a biomarker for monitoring seizure progression. doi:10.1101/2024.10.29.620811.

Chen, Z., Geng, D., & Owens, A. (2024). Images that sound: Composing images and sounds on a single canvas. Retrieved from http://arxiv.org/abs/2405.12221.

Cohen, O., Hazan, G., & Gannot, S. (2024). Multi-microphone speech emotion recognition using the hierarchical token-semantic audio transformer architecture. Retrieved from http://arxiv.org/abs/2406.03272.

Cong, G., Qi, Y., Li, L., Beheshti, A., Zhang, Z., van den Hengel, A., … Huang, Q. (2024). StyleDubber: Towards multi-scale style learning for Movie Dubbing. Retrieved from http://arxiv.org/abs/2402.12636.

Cosarinsky, M., Billot, R., Mansilla, L., Gimenez, G., Gaggión, N., Fu, G., & Ferrante, E. (2025). In-Context Reverse Classification Accuracy: Efficient estimation of segmentation quality without ground-truth. Retrieved from http://arxiv.org/abs/2503.04522.

Dixit, S., Heller, L. M., & Donahue, C. (2024). Vision language models are few-shot audio spectrogram classifiers. Retrieved from http://arxiv.org/abs/2411.12058.

Fan, C., Zhang, S., Zhang, J., Pan, Z., & Lv, Z. (2025). SSM2Mel: State Space Model to reconstruct Mel spectrogram from the EEG. Retrieved from http://arxiv.org/abs/2501.10402.

Fei, Z., Fan, M., & Huang, J. (2023). A-JEPA: Joint-Embedding Predictive Architecture can listen. Retrieved from http://arxiv.org/abs/2311.15830.

Gu, Z., Zou, H. P., Chen, Y., Liu, A., Zhang, W., & Yu, P. S. (2025). Semi-supervised in-Context Learning: A baseline study. Retrieved from http://arxiv.org/abs/2503.03062.

Guo, H., Fu, R., Geng, Y., Liu, S., Shi, S., Wang, T., … Liu, X. (2024). Mel-Refine: A plug-and-play approach to Refine Mel-spectrogram in audio generation. Retrieved from http://arxiv.org/abs/2412.08577.

Gupta, A., Likhomanenko, T., Yang, K. D., Bai, R. H., Aldeneh, Z., & Jaitly, N. (2024). Visatronic: A multimodal decoder-only model for speech synthesis. Retrieved from http://arxiv.org/abs/2411.17690.

Götz, P., Tuna, C., Brendel, A., Walther, A., & Habets, E. A. P. (2024). Blind acoustic parameter estimation through task-agnostic embeddings using latent approximations. Retrieved from http://arxiv.org/abs/2407.19989.

Hou, S., Liu, S., Yuan, R., Xue, W., Shan, Y., Zhao, M., & Zhang, C. (2024). Editing music with melody and text: Using ControlNet for Diffusion Transformer. Retrieved from http://arxiv.org/abs/2410.05151/

Huang, H., Xu, T., Xi, Y., & Chow, E. (2025). HiGP: A high-performance Python package for Gaussian Process. Retrieved from http://arxiv.org/abs/2503.02259.

Hung, Y.-N., & Lerch, A. (2022). Feature-informed embedding space regularization for audio classification. Retrieved from http://arxiv.org/abs/2206.04850.

Iakovenko, O., & Bondarenko, I. (2024). Convolutional Variational Autoencoders for Spectrogram Compression in Automatic Speech Recognition. Retrieved from http://arxiv.org/abs/2410.02560.

Iashin, V., & Rahtu, E. (2021). Taming visually guided sound generation. Retrieved from http://arxiv.org/abs/2110.08791.

Jahin, M. A., Shahriar, S., Mridha, M. F., & Dey, N. (2025). Soybean disease detection via interpretable hybrid CNN-GNN: Integrating MobileNetV2 and GraphSAGE with cross-modal attention. Retrieved from http://arxiv.org/abs/2503.01284.

Jiang, W.-B., Zhao, L.-M., & Lu, B.-L. (2024). Large Brain Model for learning generic representations with tremendous EEG data in BCI. Retrieved from http://arxiv.org/abs/2405.18765.

Kaloga, Y., Sheikh, S. A., & Kodrasi, I. (2024). Multiview Canonical Correlation Analysis for automatic pathological speech detection. Retrieved from http://arxiv.org/abs/2409.17276.

Kataria, S., Villalba, J., Moro-Velázquez, L., Żelasko, P., & Dehak, N. (2022). Time-domain speech super-resolution with GAN based modeling for telephony speaker verification. Retrieved from http://arxiv.org/abs/2209.01702.

Kim, M., Piao, Z., Lee, J., & Kang, H.-G. (2023). BrainTalker: Low-resource brain-to-speech synthesis with transfer learning using Wav2Vec 2.0. Retrieved from http://arxiv.org/abs/2312.13600.

Laouedj, S., Wang, Y., Villalba, J., Thebaud, T., Moro-Velazquez, L., & Dehak, N. (2025). Detecting neurodegenerative diseases using frame-level handwriting embeddings. Retrieved from http://arxiv.org/abs/2502.07025.

Lee, S.-H., Yoon, H.-W., Noh, H.-R., Kim, J.-H., & Lee, S.-W. (2020). Multi-SpectroGAN: High-diversity and high-fidelity spectrogram generation with adversarial style combination for speech synthesis. Retrieved from http://arxiv.org/abs/2012.07267.

Li, X., Shang, Z., Hua, H., Shi, P., Yang, C., Wang, L., & Zhang, P. (2024). SF-speech: Straightened flow for zero-shot voice clone on small-scale dataset. Retrieved from http://arxiv.org/abs/2410.12399.

Liang, Y., Liu, F., Li, A., Li, X., & Zheng, C. (2025). NaturalL2S: End-to-end high-quality multispeaker lip-to-speech synthesis with differential digital signal processing. Retrieved from http://arxiv.org/abs/2502.12002.

Lou, H., Paik, H., Haghighi, P. D., Hu, W., & Yao, L. (2024). LatentSpeech: Latent diffusion for Text-to-speech generation. Retrieved from http://arxiv.org/abs/2412.08117.

Lu, C., Zheng, W., Lian, H., Zong, Y., Tang, C., Li, S., & Zhao, Y. (2022). Speech emotion recognition via an attentive time-frequency neural network. Retrieved from http://arxiv.org/abs/2210.12430.

Mallol-Ragolta, A., Cuesta, H., Gómez, E., & Schuller, B. W. (2021). EIHW-MTG: Second DiCOVA Challenge System Report. Retrieved from http://arxiv.org/abs/2110.09239.

Nachmani, E., Levkovitch, A., Hirsch, R., Salazar, J., Asawaroengchai, C., Mariooryad, S., … Ramanovich, M. T. (2023). Spoken question answering and speech continuation using spectrogram-powered LLM. Retrieved from http://arxiv.org/abs/2305.15255.

Ouyang, B., Zhang, Y., Cheng, H., Shu, Y., Guo, C., Yang, B., … Jensen, C. S. (2025). RCRank: Multimodal ranking of Root Causes of slow queries in cloud database systems. Retrieved from http://arxiv.org/abs/2503.04252.

Pan, Y., Yang, Y., Yao, J., Ye, J., Zhou, H., Ma, L., & Zhao, J. (2024). CTEFM-VC: Zero-shot voice conversion based on content-aware Timbre Ensemble modeling and Flow Matching. Retrieved from http://arxiv.org/abs/2411.02026.

Pan, Y., Zhang, X., Yang, Y., Yao, J., Hu, Y., Ye, J., … Zhao, J. (2024). PSCodec: A series of high-fidelity low-bitrate neural speech codecs leveraging prompt encoders. Retrieved from http://arxiv.org/abs/2404.02702.

Pham, L., Lam, P., Nguyen, T., Nguyen, H., & Schindler, A. (2024). Deepfake audio detection using spectrogram-based feature and ensemble of deep learning models. Retrieved from http://arxiv.org/abs/2407.01777.

Pham, L., Lam, P., Nguyen, T., Nguyen, H., & Schindler, A. (2024). Deepfake audio detection using spectrogram-based feature and ensemble of deep learning models. Retrieved from http://arxiv.org/abs/2407.01777.

R, V. K., & Saravanan, D. (2025). A generative approach to high fidelity 3D reconstruction from text data. Retrieved from http://arxiv.org/abs/2503.03664.

Riou, A., Lattner, S., Hadjeres, G., & Peeters, G. (2024). Investigating design choices in Joint-Embedding Predictive Architectures for general audio representation learning. Retrieved from http://arxiv.org/abs/2405.08679.

Wang, J.-C., Lu, W.-T., & Won, M. (2023). Mel-band RoFormer for music source separation. Retrieved from http://arxiv.org/abs/2310.01809.

Wang, J.-C., Lu, W.-T., & Won, M. (2023). Mel-band RoFormer for music source separation. Retrieved from http://arxiv.org/abs/2310.01809.

Wu, Y., Zhang, C., Shi, J., Tang, Y., Yang, S., & Jin, Q. (2024). TokSing: Singing Voice Synthesis based on Discrete Tokens. Retrieved from http://arxiv.org/abs/2406.08416.

Yao, W., Yang, J., He, Y., Liu, J., & Wen, W. (2024). Imperceptible rhythm backdoor attacks: Exploring rhythm transformation for embedding undetectable vulnerabilities on speech recognition. Retrieved from http://arxiv.org/abs/2406.10932.

Zhan, Z., Zhou, S., Zhou, H., Liu, Z., & Zhang, R. (2025). EPEE: Towards efficient and effective foundation models in biomedicine. Retrieved from http://arxiv.org/abs/2503.02053.

Zhang, J., Zhang, T.-H., Wang, J., Gao, J., Qian, X., & Yin, X.-C. (2024). I2TTS: Image-indicated Immersive Text-to-speech synthesis with spatial perception. Retrieved from http://arxiv.org/abs/2411.13314.

Zhang, Z., Yu, Y., Chen, Y., Yang, X., & Yeo, S. Y. (2025). MedUnifier: Unifying vision-and-Language Pre-training on medical data with vision generation task using discrete visual representations. Retrieved from http://arxiv.org/abs/2503.01019.

Zhang, Z., Yu, Y., Chen, Y., Yang, X., & Yeo, S. Y. (2025). MedUnifier: Unifying vision-and-Language Pre-training on medical data with vision generation task using discrete visual representations. Retrieved from http://arxiv.org/abs/2503.01019.

Zhao, L., Chen, S., Feng, L., Zhang, X.-L., & Li, X. (2025). DualSpec: Text-to-spatial-audio generation via dual-spectrogram guided diffusion model. Retrieved from http://arxiv.org/abs/2502.18952.

Zhao, S., Pan, Z., Zhou, K., Ma, Y., Zhang, C., & Ma, B. (2025). Conditional latent diffusion-based speech enhancement via dual context learning. Retrieved from http://arxiv.org/abs/2501.10052.

Zheng, T., Wang, L., & Yu, Y. (2024). Heterogeneous space fusion and dual-dimension attention: A new paradigm for speech enhancement. Retrieved from http://arxiv.org/abs/2408.06911.

Zheng, Z., Peng, P., Ma, Z., Chen, X., Choi, E., & Harwath, D. (2024). BAT: Learning to reason about spatial sounds with large language models. Retrieved from http://arxiv.org/abs/2402.01591.

Zuo, J., Ji, S., Fang, M., Jiang, Z., Cheng, X., Yang, Q., … Zhao, Z. (2025). Enhancing expressive voice conversion with discrete pitch-conditioned flow matching model. Retrieved from http://arxiv.org/abs/2502.05471.

**Appendix A**

This appendix provides detailed information about the data generation process, model training, and architecture used in this study. The steps and mathematical formulations outlined below are essential for reproducing the results presented in the paper.

**A1. Data Generation**

The steps below outline the process of data generation.

1. Initialize Spectrogram Matrix

The spectrogram matrix is initialized as a zero matrix with dimensions $N_f \times N_t$, where:

- $N_f$ is the number of frequency bins
- $N_t$ is the number of time bins

$$S = 0_{N_f \times N_t}$$

2. Generate Chirp Signal

Linear Chirp:

The linear chirp signal is defined as:

$$S_{linear}(t) = \sin\left(2\pi \cdot \left(f_0 \cdot t + \frac{(f_1 - f_0) \cdot t^2}{2\,T}\right)\right)$$

where:

- $t \in [0, T]$ is the time vector
- $f_0$ is the starting frequency
- $f_1$ is the ending frequency
- $T$ is the duration of the chirp

Exponential Chirp:

The exponential chirp signal is defined as:

$$S_{exponential}(t) = \sin\left(2\pi \cdot f_0 \cdot \frac{(k^t - 1)}{\ln(k)}\right)$$

where:

- $k = \left(\frac{f_1}{f_0}\right)^{\frac{1}{T}}$ is the exponential growth factor.
- $f_0, f_1, t$ and $T$ are as defined above.

3. Map Chirp Signal to Spectrogram Matrix

Time Bins:

The start and end time bins for the chirp are calculated as:

$$t_{start} = \left[\frac{t_0}{T} \cdot N_t\right]$$

$$t_{end} = \left[\frac{t_0 + \Delta t}{T} \cdot N_t\right]$$

where:

- $t_0$ is the start time of the chirp
- $\Delta t$ is the duration of the chirp

Frequency Bins:

The start and end frequency bins for the chirp are calculated as:

$$f_{start} = \left[\frac{f_0}{f_{max}} \cdot N_f\right]$$

$$f_{end} = \left\lfloor \frac{f_1}{f_{max}} \cdot N_f \right\rfloor$$

where:

- $f_{max}$ is the maximum frequency.

Mapping the Chirp to the Spectrogram:

For each time bin $t_b \in [t_{start}, t_{end}]$:

1- Compute the normalized time:

$$\tau = \frac{t_b - t_{start}}{t_{end} - t_{start}}$$

2- Compute the frequency bin:

- Linear Chirp:

$$f_b = f_{start} + (f_{end} - f_{start}) \cdot \tau$$

- Exponential Chirp:

$$f_b = f_{start} \cdot \left(\frac{f_{end}}{f_{start}}\right)^\tau$$

3- Spread the intensity around the frequency bin using a Gaussian weight:

$$\omega(f) = exp\left(-\frac{(f - f_b)^2}{2\sigma^2}\right)$$

where:

- $f$ is the frequency bin index.
- $\sigma$ controls the spread of the Gaussian (e.g., $\sigma = 0.5$)

4. Add Gaussian Noise and Apply Gaussian Filter

Add Gaussian Noise:

Gaussian noise is added to the spectrogram:

$$S_{noisy} = S + \eta \cdot N$$

where:

- $\eta$ is the noise level, sampled uniformly from $[0.09, 0.3]$
- N is a matrix of random values sampled from a standard normal distribution: $N \sim \mathcal{N}(0,1)$

Apply Gaussian Filter:

A Gaussian filter is applied to smooth the spectrogram:

$$S_{smooth} = G_\sigma * S_{noisy}$$

where:

- $G_\sigma$ is a 2D Gaussian kernel with standard deviation $\sigma$ ($e.g.$, $\sigma = 1$)
- $*$ denotes the convolution operation.

5. Save Spectrogram and Labels

The spectrogram is saved as an image:

$$image\_path = os.path.join(output\_dir, f"spectrogram_t.png")$$

The chirp parameters are saved in a CSV file:

$$labels\_file = os.path.join(output\_dir, "labels.csv")$$

Each row in the CSV file contains:

$t_0, f_0, f_1, \Delta t, chirp\_type$

## A2. Model training

Model training involves several critical steps to ensure the model learns effectively from the data and generalizes well to unseen examples. The process begins with Label Normalization, where the labels are standardized to have zero mean and unit variance. This step ensures that all features contribute equally to the learning process, preventing any single feature from dominating due to differences in scale. Next, the Dataset Splitting step divides the data into training and testing sets, allowing for proper evaluation of the model's performance on unseen data. This is followed by Image Preprocessing, where images are resized, normalized, and transformed into tensors to match the input requirements of the model. These preprocessing steps are essential for preparing the data in a format suitable for training deep learning models. Finally, the model is trained using the preprocessed data, with the normalized labels serving as the target outputs. This structured approach ensures that the model is trained on standardized data.

**Label Normalization**

Let $Y \in R^{N \times 3}$ be the matrix of labels, where $N$ is the number of samples and each row corresponds to the three features: Chirp Start Time, Chirp Start Freq, and Chirp End Freq.

The mean $\mu \in R^3$ and standard deviation $\sigma \in R^3$ are computed as:

$$\mu = \frac{1}{N} \sum_{i=1}^{N} Y_i$$

$$\sigma = \sqrt{\frac{1}{N} \sum_{i=1}^{N} (Y_i - \mu)^2}$$

The labels are normalised as:

$$Y_{norm} = \frac{Y - \mu}{\sigma}$$

**Dataset Splitting**

The dataset is split into training and testing sets using a random partition:

$$\mathcal{T}_{train}, \mathcal{T}_{test} = Split(\mathcal{T}, test_{size} = 0.2, random_{state} = 42)$$

where:

- $\mathcal{T} = \{1, 2, \ldots, N\}$ is the set of indices
- $\mathcal{T}_{train}$ and $\mathcal{T}_{test}$ are the training and testing indices, respectively

**Image Preprocessing**

For each image $X_i \in R^{H \times W \times C}$, where $H$ is height, $W$ is width, and $C$ is the number of channels (3 for RGB), the preprocessing steps are:

Resize the image $224 \times 224$:

$$X_i' = Resize(X_i, (224, 224))$$

Normalize pixel values to $[0,1]$:

$$X_i'' = \frac{X_i'}{255}$$

Convert to a tensor and permute dimensions:

$$X_i''' = Permute(X_i'', (2, 0, 1))$$

**A3. Vision Transformer (ViT) Model**

The Vision Transformer (ViT) model is a state-of-the-art architecture that leverages the transformer mechanism to process images. The model consists of three main components: Patch Embedding, where the image is split into fixed-size patches and projected into a lower-dimensional space; Transformer Encoder, which applies self-attention mechanisms to capture global relationships between patches; and a Regression Head, which predicts the target outputs using the CLS token representation. The model is trained using Mean Squared Error (MSE) loss, optimized with AdamW, and employs a learning rate scheduler and early stopping to improve convergence and prevent overfitting. Evaluation metrics such as Pearson correlation and inference speed are used to assess performance, and predicted outputs are denormalized to their original scale for interpretation. This approach enables the ViT model to learn spatial and contextual features from images for regression tasks.

The ViT model processes the input image $X_i'''$ as follows:

**a. Patch Embedding**

The image is divided into patches of size $16 \times 16$, and each patch is flattened into a vector:

$$P = PatchEmbed(X_i''')$$

where $P \in \mathbb{R}^{M \times D}$, $M$ is the number of patches, and $D$ is the embedding dimension.

**b. Transformer Encoder**

The patches are passed through a transformer encoder:

$$H = TransformerEncoder(P)$$

where $H \in \mathbb{R}^{M \times D}$ is the hidden state representation.

**c. Regression Head**

The [CLS] token representation $h_{cls} \in \mathbb{R}^D$ (first row of H) is passed through a regression head:

$$y_{pred} = W_2 \cdot ReLU(W_1 \cdot h_{cls} + b_1) + b_2$$

where:

$$W_1 \in \mathbb{R}^{256 \times D}, b_1 \in \mathbb{R}^{256}$$
$$W_2 \in \mathbb{R}^{128 \times 256}, b_2 \in \mathbb{R}^{128}$$
$$W_3 \in \mathbb{R}^{3 \times 128}, b_3 \in \mathbb{R}^{3}$$

**Loss Function**

The Mean Squared Error (MSE) loss is computed as:

$$L = \frac{1}{B} \sum_{i=1}^{B} \left\| y_{pred}^{(i)} - y_{true}^{(i)} \right\|_2^2$$

where:

$B$ is the batch size.

$y_{pred}^{(i)}$ is the predicted output for the $i_{th}$ sample.

$y_{true}^{(i)}$ is the true label for the $i_{th}$ sample.

**Optimization**

The AdamW optimizer updates the model parameters θ as:

$$\theta_{t+1} = \theta_t - \eta \cdot AdamW(\nabla_\theta L, \theta_t)$$

where:

$\eta$ is the learning rate.

$\nabla_\theta L$ is the gradient of the loss with respect to the parameters.

**Learning Rate Scheduler**

The learning rate is adjusted based on the validation loss:

$$\eta_{t+1} = \begin{cases} \eta_t \cdot \gamma & \text{if } L_{val} \text{ does not improve for } k \text{ epochs,} \\ \eta_t & \text{otherwise} \end{cases}$$

where:

- $\gamma$ is the factor by which the learning rate is reduced.
- $k$ is the patience parameter.

**Evaluation Metrics**

**Pearson Correlation**

The Pearson correlation coefficient $r$ between predicted and true labels is computed as:

$$r = \frac{cov(y_{pred}, y_{true})}{\sigma_{y_{pred}} \cdot \sigma_{y_{true}}}$$

**Inference Speed**

The inference time $t_{inf}$ is measured as:

$$t_{inf} = t_{end} - t_{start}$$

**Denormalization**

Predicted values are denormalized as:

$$y_{pred}^{denorm} = y_{pred} \cdot \sigma + \mu$$

**Early Stopping**

Training stops if the validation loss does not improve for $p$ epochs:

$$\text{Stop if } L_{val}^{(t)} \geq L_{val}^{(t-p)} \text{ for } p \text{ consecutive epochs.}$$

**A4. The model architecture**

Vision Transformer (ViT) was adapted for the regression task, with LoRA (Low-Rank Adaptation) fine-tuning applied to the attention layers.

**Input Embedding (ViT Embeddings):**

- Input: Image $X \in \mathbb{R}^{H \times W \times 3}$
- Patch Embedding:
    - Convolutional projection: Conv2d(3,768, kernel size = (16,16), stride = (16,16))
    - Output: $Z \in \mathbb{R}^{N \times 768}$, where $N = \frac{H \times W}{16 \times 16}$ is the number of patches.
- Dropout: $Z' = Dropout(Z, p = 0.0)$

**Transformer Encoder (ViT Encoder):**

The encoder consists of $L = 12$ layers, each with the following structure.

Self-Attention (ViT SDPA Attention)

- $Query\ (Q), Key\ (K), and\ Value\ (V)\ projections$:
    - $Q = LoRA(Linear(Z'))$, $where\ LoRA\ introduces\ low\ rank\ adaptation$:
        - $Q = W_Q Z' + \Delta W_Q Z'$, $with\ \Delta W_Q = B_Q A_Q, A_Q \in \mathbb{R}^{768 \times 8}, B_Q \in \mathbb{R}^{8 \times 768}$
    - $K = Linear(Z')$
    - $V = LoRA(Linear(Z')), similar\ to\ Q$
- $Scaled\ DotProduct\ Attention$:
    - $Attention(Q, K, V) = Softmax\left(\frac{QK^T}{\sqrt{d_k}}\right) V$, $where\ d_k = 768$
- $Output\ projection: Linear(Attention(Q, K, V))$

Feed-Forward Network (FFN):

- $Intermediate\ layer: Linear(768 \rightarrow 3072), followed\ by\ GELU\ activation$
- $Output\ layer: Linear(3072 \rightarrow 768)$

Layer Normalization:

- Applied before and after the self-attention and FFN blocks:
    - $LayerNorm(z')$

Pooling (ViT Pooler):

- $Pooled\ representation: Tanh(Linear(Z')), where\ Z'\ is\ the\ output\ of\ the\ encoder$

Regression Head:

- Sequential fully connected layers:
    1. $Linear(768 \rightarrow 256), followed\ by\ ReLU$
    2. $Linear(256 \rightarrow 128), followed\ by\ ReLU$
    3. $Linear(128 \rightarrow 3), producing\ the\ final\ regression\ output$